\documentclass[11pt]{article}
\pdfoutput=1
\usepackage[a4paper, total={6.5in, 8in}]{geometry}

\usepackage{setspace} 
\usepackage[utf8]{inputenc} 
\usepackage{graphicx}
\usepackage{bm}

\usepackage[english]{babel}
\usepackage{ulem}

\usepackage{xcolor}
\newcommand{\rf}[1]{(\ref{#1})}

\newcommand{\beq}{\begin{equation}}
\newcommand{\eeq}{\end{equation}}
\newcommand{\beqr}{\begin{eqnarray}}
\newcommand{\eeqr}{\end{eqnarray}}

\newcommand{\lb}[1]{\label{#1}}

\newcommand{\bc}{\begin{center}}
\newcommand{\ec}{\end{center}}
\newcommand{\ct}[1]{\cite{#1}}

\begin{document}
\bc{\bf\large  Surface and volume photoemission from metal\\ nano-particles with electron mass discontinuity}\\{\itshape I.E.Protsenko and A.V.Uskov}\\
Quantum Electronic division, Lebedev Physical Institute
Moscow, Russia 119991\\ \today

\parbox{14cm}{\begin{singlespace}
{\small\bc {\bf Abstract}\ec Quantum efficiencies of surface (SPE) and volume  (VPE) photo-emissions from metal nanoparticles are calculated by quantum mechanical perturbation theory and compared with each other. Along with discontinuities in the potential barrier and dielectric function, the discontinuity in electron effective mass on the metal-environment interface is taken into account. General formulas for quantum efficiencies of SPE and VPE are derived. An example of spherical gold particles with rectangular potential barrier on  the interface is considered, analytical formulas for quantum efficiencies of SPE and VPE on the red border of photoemission are derived.   It is found that the efficiency of SPE is less decreased with the reduction of the electron effective mass than the efficiency of VPE, so SPE is more efficient that VPE for small particles and large discontinuity in effective mass. Nanoparticle size, when SPE is more efficient than VPE, is found to be tens of nm or less. }\end{singlespace}}\ec


\section{Introduction}
A photon of electromagnetic field in a metal can be absorbed  at the collision of an electron with impurity, lattice defect  or with the metal surface \ct{Kittel2004}. If the energy of an electron, after absorption of a photon, exceeds the potential barrier height on the metal-environment interface, the electron, with some probability, can be emitted from the metal.  In  metal nano-particles, at certain conditions, high-energy electrons, generated by the  photo-absorption, become "hot": they  form an ensemble  characterized by a model carrier density and by a distribution with some high temperature  \ct{doi:10.1002/adom.201901166}. Hot electrons find various applications \ct{article_3}, so the increase of efficiency of hot electron generation is a topical  problem. It is interesting to use localized plasmon resonances (LPR) with large energy densities of the field in metal nanostructures for increasing the efficiency of hot electron generation. However LPR frequencies, typically, are in the visible or IR regions, so such field quanta can't overcome potential barriers in the metal-dielectric enviroment, and the photoemission is forbidden.
In order to overcome this difficulty and utilize LPR for the photoemossion one can insert metal nanoparticle into a semiconductor, and reduce, this way, the work function of electron in metal \ct{article_3}. However, the electron emission from a metal into a semiconductor environment  is strongly reduced, because of the electron effective mass decreases, while an electron goes from a metal to a semiconductor, so an electron must be additionally accelerated in the requirement of the momentum conservation law \ct{Khurgin}. 

Photoemission from the surface of nanoparticles \ct{C3NR06679G} gives a possibility to increase the hot electron generation,  despite of the problem with the electron mass decrease, which has been confirmed by recent experiments \ct{doi:10.1021/acscatal.9b00384}. The purpose of this paper is to justify theoretically, that  surface photoemission (SPE) is, indeed, less reduced due to the change in the electron mass on the metal-semiconductor interface that the photoemission from the metal volume. Such an advantage of SPE is added to other advantages of SPE respectively to volume photoemission (VPE) discussed in \ct{Protsenko_2012, C3NR06679G}.   
 
 Absorption of electromagnetic field at collisions of electrons with the metal-environment  interface and emission of  hot electrons outside the metal have been studied for a long time \ct{Tamm}. Two processes are possible at the absorption of a photon by an electron in the collision with the surface  \ct{Brodsky}. One process is a surface photo-emission (SPE), another process is when an  electron absorbs a photon and came back to a metal, this is surface photo-absorption (SPA).  Discontinuities  in the potential barrier, electromagnetic field and  electron effective mass on the interface strongly influence SPE and SPA  \ct{Brodsky, Brodsky2}. General theoretical mpdel of SPE in metal nanoparticles \ct{Protsenko_2012}    follows the description of SPE  by quantum theory of  perturbations in continues spectrum \ct{Brodsky,Brodsky2}, and takes into account all these  discontinuities. The approach and results of \ct{Protsenko_2012} have been used, in particular, for studies of electron photoemission
from plasmonic nanoantennas  \ct{doi:10.1002/pip.2278},  comparison between the surface and the
volume photoelectric effects in  photoemission from plasmonic
nanoparticles \ct{C3NR06679G},  broadening of plasmonic resonance due to electron collisions with nanoparticle boundary \ct{Uskov2014} and enhanced electron photoemission by collective resonances in the nanopartical lattices \ct{Zhukovsky2014}. 

However, the comparison between the surface and the volume photoemission in metal nanoparticles, taking into account the discontinuity on the electron effective mass, is not yet made. A comparison of SPE and VPE in \ct{C3NR06679G} has been made without  taking account a change (a discontinuity) in the electron effective mass on the metal-semiconductor interface.   In this paper we continue the study of \ct{Protsenko_2012, C3NR06679G} and compare SPE and VPE taking into account the discontinuity of electron effective mass along with another discontinuities in the metal-semiconductor interface. Such comparison will help to find conditions for the most efficient emission of hot electrons from metal nanoparticles using both VPE and SPE. Here we also expand results of  \ct{Protsenko_2012}  to the case of  SPA and provide some general formulas for quantum efficiencies of SPE and SPA derived all together, in commonly used approximations. 
SPA is important as a source of additional broadening of surface plasmon resonances in nano-particles \ct{Ikhsanov:20}, for correct estimations of the heating of metal nano-structures necessary, for example, for  medical applications \ct{doi:10.1002/advs.201900471}.    

In Section~\ref{Sec2} we, following \ct{Protsenko_2012}, consider the surface photoemission and photoabsorption, provide general formulas for internal quantum efficiencies of SPE and SPA, consider SPE and SPA with restangular-step potential barrier and SPE near the red border of photoemission. As a starting point, we present results for auxiliary one-dimentional (1D) problem and then consider more general 3D problem. We estimate the maximum size of nanoparticles, when SPE became important and has the same efficiency that VPE. 

In Section~\ref{Sec3} we, following \ct{C3NR06679G}, consider the volume photoemission, provide general formula for its internal quantum efficiency, consider VPE for rectangular-step potential barrier, in particular, near the red border. 

In Section~\ref{Sec4} we compare efficiencies of the volume and the surface photoemissions, explain and discuss the results of such comparison. Through the paper we use examples with spherical gold nanoparticle in various semiconductor environments and with rectangular-step potential barrier on the border. We   do not provide detailed derivation of formulas, but describe them in all details sufficient for the application. We summarize results in the Summary section.

\section{Surface photoemission}\label{Sec2}
\begin{figure}[ht]
\bc \includegraphics[width=8.6cm]{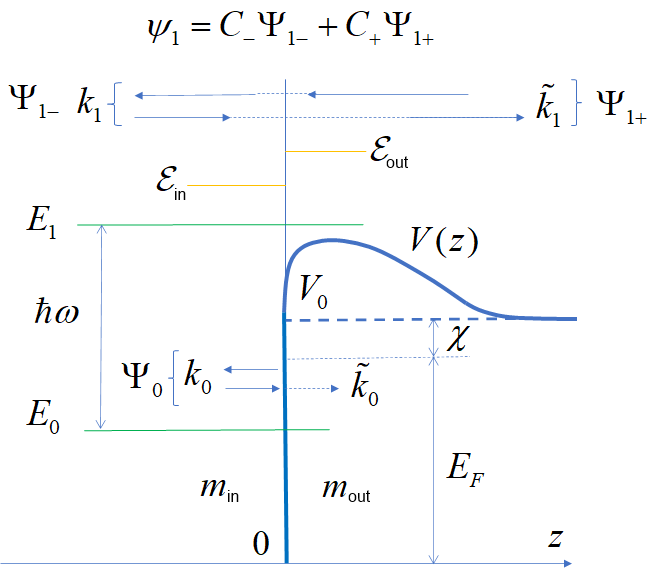} \ec
\vspace{-4mm}
\caption{Electron from metal moves toward and collides with potential barrier $V(z)$ on metal ($z<0$) - semiconductor ($z>0$) interface in $z=0$. Colliding with the barrier, the electron absorbs a photon of energy $\hbar\omega$ from external  electromagnetic field. $V(z)$ has a step change from $0$ to $V_0$ in $z=0$, $V\rightarrow V_0$ at $z\rightarrow +\infty$, $V_0$ is the energy of the bottom of  conductive band of a semiconductor. $V_0 = E_F+\chi$, $E_F$ is Fermi-energy and $\chi$ is a work function. Medium dielectric function $\varepsilon(z)$ and electron effective mass $m(z)$ have step changes  in $z=0$ between values $\varepsilon_{out}$, $m_{out}$ for $z>0$ in semiconductor and $\varepsilon_{in}$, $m_{in}$ for $z<0$ in metal. $E_0$ ($E_1$) are energies of the electron in initial (final) states of absorption. $\Psi_0$ is initial wave-function of the electron with the wave-number $k_0$ ($\tilde{k}_0$) at $z<0$ ($z>0$). $\Psi_{1+}$ and $\Psi_{1-}$ are  wave-functions of electron in the basic final states with wave-numbers $k_1$ at $z<0$ and $\tilde{k}_1$ at $z\rightarrow +\infty$,  $\Psi_{0,1\pm}$ are calculated without electromagnetic field. $\psi_1$ is the wave-function of final state of an electron after absorption of a photon calculated by perturbation theory. }\label{Fig1}
\end{figure}

Suppose an electron in a metal moves toward the interface between the metal and the semiconductor in external monochromatic electromagnetic field of  frequency $\omega$. Following the approach of \ct{Brodsky2} we consider, first, one-dimension (1D) problem, when the interface is described by 1D potential barrier $V(z)$ shown on Fig.\ref{Fig1}. The electron reaches the interface,  collides with  the barrier and  absorbs, with some probability, a photon from the external field. Only the component of external field normal to the  interface interacts with the electron. The electron may pass  the barrier, leave the metal, and contribute to the surface photo-emission (SPE). Probability amplitude of such process is $C_+$. Otherwise, the electron may be reflected from the barrier back to the metal with a probability amplitude $C_-$ and contribute to the surface photo-absorption (SPA).  
%
%
%
%
%
\subsection{Results for 1D case}
The interaction of an electron with the electromagnetic field is a perturbation. Final state $\psi_1$ of an electron after absorption of a photon is $\psi_1 = C_+\Psi_{1+}+C_-\Psi_{1-}$, where $\Psi_{1\pm}$ are unperturbed wave-functions of the electron moving in the potential $V(z)$ without electromagnetic field. We found probability amplitudes of SPE $C_+$ and SPA $C_-$  
\beq
C_{\pm} = c_{\pm}\mathcal{E}_{in}^{(n)} = \left(c_{\pm}^{(0)} -
\frac{|e|\varepsilon_{in}}{W(\hbar\omega)^2\varepsilon_{out}}\int_{+0}^{\infty} V'\Psi_0\Psi_{1\mp}dz\right)\mathcal{E}_{in}^{(n)},\lb{New1}
\eeq
where $\mathcal{E}_{in}^{(n)}$ is the amplitude of the component of electric field inside the metal, on the interface and  normal to the interface. Coefficient $c_{\pm}^{(0)}$ describe the break in the potential $V$, dielectric function $\varepsilon$ and electron effective mass $m$ on the interface
\beq
c_{\pm}^{(0)} = \frac{|e|}{(\hbar\omega)^2}\frac{m}{m_{in}W}\left\{\left[\left(r_{\varepsilon}r_m - 1\right)\left(E_0+\frac{\hbar\omega}{2}\right) -
r_{\varepsilon}r_mV_0\right]\Psi_0\Psi_{1\mp}
+(r_{\varepsilon}-1)m_{in}\left.\frac{\hbar^2\Psi_0'\Psi_{1\mp}'}{2m^2}\right\}\right|_{z=0},\lb{2}
\eeq
where $r_{\varepsilon} = \varepsilon_{in}/\varepsilon_{out}$, $r_m = m_{in}/m_{out}$ and $\Psi_0'\Psi_{1\mp}'/{m^2}$ is continues function of $z$ also in $z=0$. Expression \rf{2}  follows from Eq.(2) of \ct{Protsenko_2012} as shown in Appendix. The second term in Eq.\rf{New1}  describes the interaction of an electron with potential barrier $V(z)$ at $z>0$. We suppose that discontinuities in $V$, $\varepsilon$ and $m$ are in $z=0$; $m$ and $\varepsilon$ at $z>0$ and $z<0$ are constant, $V(z)$ is continues function of $z$ for $z>0$. Parameters used in Eq.\rf{2} are shown and described in the caption of Fig.\ref{Fig1}.

Wave-functions $\Psi_0$, $\Psi_{1\pm}$ are solutions of the "unperturbed" Schrodinger equation
\beq
        \left[-\frac{\hbar^2}{2}\frac{d}{dz}\left(\frac{1}{m}\frac{d}{dz}\right) + V(z)\right]\Psi(z) = E\Psi(z). \lb{ShEq}
\eeq
without the interaction with electromagnetic field. The first term on the left in Eq.\rf{ShEq} is kinetic energy operator of electron with effective mass $m(z)$. In Eq.~\rf{2} $\Psi_0$ is the wave function of the electron in  initial state, $\Psi_0(z<0) = e^{ik_0z}$ with the wave number $k_0=\sqrt{2m_{in}E_0}/\hbar$ and the energy $E=E_0$. 
The final state of the electron after absorption of a photon is 
\[
    \psi_1 = C_+(z)\Psi_{1+}(z)+C_-(z)\Psi_{1-}(z),
\]
where $\Psi_{1+}$ ($\Psi_{1-}$) describe the electron with energy $E_1 = E_0+\hbar\omega$. Far from the interface such electron is moving  to the right (to the left).  $C_+$  is the probability amplitude of SPE. Wave function $\Psi_{1+}$ is normalized such that  $\Psi_{1+}(z\rightarrow\infty)\rightarrow e^{i\tilde{k}_1z}$ describes the electron of photo-emission current with the wave number $\tilde{k}_{1}=\sqrt{2m_{out}(E_1-V_0)}/\hbar$ moving to the right in conduction band of the semiconductor environment. $C_-$ is the probability amplitude of SPA, $\Psi_{1-}$ is normalized such that $\Psi_{1-}(z<0)= e^{-ik_1z}$ corresponds to the electron with the wave number $k_1=\sqrt{2m_{in}E_1}/\hbar$ reflected from the barrier and moving to the left in the metal, such electron contributes to SPA. Wronskian
\[W(\Psi_{1-}\Psi_{1+}) = \Psi_{1-}\frac{d\Psi_{1+}}{dz} - \Psi_{1+}\frac{d\Psi_{1-}}{dz},\]
$W/m$ does not depend on $z$ \ct{Protsenko_2012}. In Eq.\rf{2}  prime means the derivative over $z$. 

\subsection{Results for 3D case}
In $3D$ case an electron moves at arbitrary angle to the metal-semiconductor interface and the electromagnetic field is not necessary perpendicular to the interface. We  factorize 3D wave function of an electron  
\beq
\Psi_i(z)e^{i\vec{k}_{\parallel}\vec{\rho}}, \lb{Fac_wf}
\eeq
where indexes $i=\{0,1+,1-\}$ have the same meaning as in 1D case, $e^{i\vec{k}_{\parallel}\vec{\rho}}$ describes the motion of an electron parallel to the interface,  $\vec{k}_{\parallel} = \{ \vec{e}_{x}{k}_{x}, \vec{e}_{y}{k}_{y} \}$, $\vec{\varrho} = \{ \vec{e}_{x}x, \vec{e}_{y}y \}$ and $\vec{e}_{x}$, $\vec{e}_{y}$ are unit vectors of Cartesian coordinate system with axes $x,y$ parallel to the interface. Approximation \rf{Fac_wf} has been used in \ct{Brodsky, Brodsky2}  supposing that potential barrier is averaged over $x$ and $y$ and "flat" in these directions. In such "flat" approximation for $V(z)$, medium dielectric function $\varepsilon$ and electron effective mass $m$ do not depend on $x$ and $y$, so $\vec{k}_{\parallel}$ in Eq.~\rf{Fac_wf} are preserved and remain the same as in the initial state of an electron.

Multipliers $\Psi_{0,1\pm}(z)$ in Eq.~\rf{Fac_wf} are solutions of Eq.~\rf{ShEq}, which  describes  the motion of an electron perpendicular to the interface. In Eq.~\rf{ShEq} initial energy $E=E_z = (\hbar k_z)^2/2m_{in}$, which is a part of kinetic energy related of the motion of electron in metal perpendicular to the interface,  $k_z$ is z-component of the electron wave vector. Only the component of electromagnetic field perpendicular to the interface interacts with the electron at the collision, so the energy of a photon, absorbed by electron, increase the part of kinetic energy of an electron correspondent to the motion perpendicular to the interface. A part of kinetic energy of an electron in metal, in final state, related with the motion perpendicular to the interface, is $E_{1z} = E_z+\hbar\omega$.  

The second term in Eq.~\rf{New1} is the same in 1D and in 3D cases. In Eq.\rf{2}  $E_0$   must be replaced by $E_z=(\hbar k_z)^2/2m_{in}$ and $V_0$ is replaced by
\beq
   V_z=V_0+\frac{\hbar^2k_{\parallel}^2}{2}\left(\frac{1}{m_{out}}-\frac{1}{m_{in}}\right). \lb{new_10}
\eeq
Replacement \rf{new_10} is a consequence of  the momentum conservation law, which preserves $k_{\parallel}$, while an electron passes through the "flat" interface with the discontinuity of the effective mass. 
Typically $m_{in}>m_{out}$ in the metal-semiconductor interface, so the momentum conservation law requires increase of the electron velocity in the semiconductor and, therefore,  the electron kinetic energy, which leads to increase of potential barrier $V_z>V_0$, as it is in Eq.~\rf{new_10}. 

Only electrons with  
$E_z + \hbar\omega > V_z$ 
pass the barrier on the interface and contribute to SPE.  $z$-component of velocity of such electrons far from the barrier is $\hbar\tilde{k}_{1z}/m_{out}$ where 
\beq
\tilde{k}_{1z} = \sqrt{[k_z^2+k_{\omega}^2 - k_V^2-k_{\parallel}^2(r_m-1)]/r_m}, \hspace{0.5cm} k_{\omega} = \sqrt{{2m_{in}\omega}/{\hbar}}, \hspace{0.5cm} k_V = {\sqrt{2m_{in}V_0}}/{\hbar}. \lb{k1z}
\eeq
Now we write expressions for SPE and SPA currents in electrons per second. We consider "cold" electrons with initial energies in small interval $E_0\div E_0+dE_0$. A part of such electrons participate in SPE and SPA and make photo-currents with surface densities 
\beq
dJ_{SPE,SPA} =|\mathcal{E}_{in}^{(n)}|^2 dj_{SPE,SPA}, \lb{dif_surf_cur}
\eeq
where
\beq
        dj_{SPE} = \frac{\hbar\Re{(\tilde{k}_{1z}})}{m_{out}}|c_+|^{2}dn_s,  \hspace{0.5cm}
 dj_{SPA} = \frac{\hbar{k}_{1z}}{m_{in}}|c_-|^2dn_s, \lb{dj}
\eeq
are photo-current surface densities per unit of $|\mathcal{E}_{in}^{(n)}|^2$, $\mathcal{E}_{in}^{(n)}$ is  the component  of the field inside metal normal to the interface,
\beq
    dn_s = f_F(k_0^2)[1-f_F(k_0^2+k_{\omega}^2)]\frac{2k_{\parallel}dk_{\parallel}dk_z}{(2\pi)^2} \lb{dne}
\eeq
is a number of "cold" electrons, which can absorb a photon, in the unit of volume in small interval of energies $f_F(k_0^2)[1-f_F(k_0^2+k_{\omega}^2)]$ is the probability that the initial state of an electron with the energy $E_0 = (\hbar k_0)^2/2m_{in}$ is occupied and the final state with the energy $E_1 = \hbar^2 (k_0^2+k_{\omega}^2)/2m_{in}$ is free;  $k_0^2=k_z^2+k_{\parallel}^2$, $k_{1z} = \sqrt{k_z^2+k_{\omega}^2}$. In Eq.\rf{dj}  we take real part of $\tilde{k}_{1z}$  because of SPE exists only for electrons with a part of kinetic energy, related with the motion along axes $z$, above the bottom energy of semiconductor conductive band, so when  
\beq
k_z^2+k_{\omega}^2 - k_V^2-k_{\parallel}^2(r_m-1)>0, \lb{cond_Re_k}
\eeq
and $\tilde{k}_{1z}$ is real. In Eq.~\rf{dne} 
\beq
        f_F(y) = \left[1+\exp{\frac{\hbar ^2y/2m_{in} - E_F}{K_BT}}\right]^{-1} \lb{Fermi_df}
\eeq
is Fermi-distribution function, $K_B$ is  Boltzmann constant  and $T$ is a temperature.

In order to obtain total photo-currents $J_{SPE}$ of SPE and $J_{SPA}$ of SPA  we integrate  photo-current densities 
$dJ_{SPE,SPA}$ \rf{dif_surf_cur} over all states of electrons colliding with nano-particle surface, i.e. over $dk_{\parallel}$, $dk_z$, and over the nanoparticle surface with the area $S_p$. In the flat approximation $dj_{SPE,SPA}$ are the same in any point of the surface, so integrations over electrons and over the surface are independent on each other and
\beq
        J_{SPE,SPA} = \int_{k_z,k_{\parallel}}  dj_{SPE,SPA}\int_{S_p} |\mathcal{E}_{in}^{(n)}|^2 dS_p.    \lb{J_SPESPA}
\eeq
Internal quantum efficiency of SPE or SPA is
\beq
        \eta_{SPE,SPA} = J_{SPE,SPA}/R_{abs}, \lb{Int_q_eff}
\eeq
where 
\beq
R_{abs}=\int_{V_p} r_{abs}dV \lb{RABS}
\eeq
is the rate of absorption of photons in the volume $V_p$ of a nanopatricle,
\beq
r_{abs}=\frac{\varepsilon_{in}''}{2\pi\hbar}|\mathcal{E}_{in}|^2 \lb{rabs_v}
\eeq
is the absorption rate in the unit of volume, $\varepsilon_{in}''$ is imaginary part of the dielectric function of metal.  Combining expressions \rf{J_SPESPA} -- \rf{rabs_v} we find
\beq
        \eta_{SPE,SPA} = \frac{2\pi\hbar}{\varepsilon_{in}''a}\int_{k_z,k_{\parallel}}dj_{SPE,SPA}, \lb{eta_1}
\eeq
where the length  
\beq
        a = \int_{V_p}|\mathcal{E}_{in}|^2dV_p/\int_{S_p} |\mathcal{E}_{in}^{(n)}|^2 dS_p. \lb{geom_fac}
\eeq
In general $a$ depends on the field.  For  small spherical nanoparticles with uniform electric field inside, $a$ is nanoparticle radius.  

It is convenient to introduce dimensionless variables
\beq
x_z = (k_z/k_V)^2, \hspace{0.5cm} x_{\parallel} = (k_{\parallel}/k_V)^2, \hspace{0.5cm} x_{\omega} = (k_{\omega}/k_V)^2 = \hbar\omega/V_0, \hspace{0.5cm} x_F=E_F/V_0 \lb{dim_en}
\eeq
and write
\beq
\eta_{SPE} = \frac{\eta_s}{\sqrt{r_m}}\int\Re{\sqrt{x_z+x_{\omega} - 1-x_{\parallel}(r_m-1)}}|\tilde{c}_+|^2d\tilde{n}_s,\hspace{0.5cm}
\eta_{SPA} = \eta_s\int\sqrt{x_z+x_{\omega}}|\tilde{c}_-|^2d\tilde{n}_s. \lb{eta_SPE_SPA_nodim}
\eeq
An electron passes the barrier, if $\eta_{SPE}$ is real, so  when
\beq
x_z+x_{\omega} - 1-x_{\parallel}(r_m-1)>0, \lb{SPE_cond}
\eeq
this is why the real part is taken  in the first one of Eqs.~\rf{eta_SPE_SPA_nodim}. There
\beq
        \eta_s = \frac{\lambda/a}{(2\pi)^2\varepsilon_{in}''}\frac{e^2}{\hbar  c}\left(\frac{V_0}{\hbar\omega}\right)^3, \hspace{0.5cm} d\tilde{n}_s = \tilde{f}_F(x)[1-\tilde{f}_F(x+x_{\omega})]\frac{dx_{\parallel}dx_z}{\sqrt{x_z}}, \lb{F_tilde}
\eeq
where $\lambda$ is wavelength of the external field in vacuum, $e^2/\hbar c \approx 1/137$ is fine structure constant. For gold nanoparticle in p-doped silica, considered in \ct{Protsenko_2012}, $\eta_s = 1$ for $a =  15$~nm near localized plasmon resonance $\lambda = 0.857$~$\mu$m.  Fermi distribution function
\beq
        \tilde{f}_F(x) = \left[1+\exp{\left(\frac{x-x_F}{x_T}\right)}\right]^{-1}, \hspace{0.5cm} x_F=E_F/V_0, \hspace{0.5cm}  \lb{Fermi_df_1}
\eeq
and
\beq
     \tilde{c}_{\pm} =  \tilde{c}_{\pm}^{(0)} -
\frac{\varepsilon_{in}k_V}{V_0W\varepsilon_{out}}\int_{+0}^{\infty} V'\Psi_0\Psi_{1\mp}dz,  \lb{cpm_til}
\eeq\beq
     \tilde{c}_{\pm}^{(0)} =  \frac{mk_V}{m_{in}W}\left\{\left[\left(r_{\varepsilon}r_m - 1\right)\left(x_z+\frac{x_{\omega}}{2}\right) -
r_{\varepsilon}r_m\left[1+x_{\parallel}\left(r_m-1\right)\right]\right]\Psi_0\Psi_{1\mp}
+(r_{\varepsilon}-1)m_{in}\left.\frac{\hbar^2\Psi_0'\Psi_{1\mp}'}{2m^2V_0}\right\}\right|_{z=0}. \lb{cpm_til_c0}
\eeq
When we write Eq.~\rf{cpm_til_c0} we replace in Eq.~\rf{2} $E_0$  by $E_z=(\hbar k_z)^2/2m_{in}$ and $V_0$ by $V_z$ given by Eq.~\rf{new_10}. 

SPE is increased, when a nanoparticle became smaller. We show the dependence on the nanoparticle size explicitly and represent  
\beq
        \eta_s = a_s/a, \hspace{0.5cm} a_s(x_{\omega})= a_{s0}\frac{\varepsilon_{rb}''}{\varepsilon''(x_{\omega})}\left(\frac{1-x_F}{x_{\omega}}\right)^4, \hspace{0.5cm} a_{s0} = \frac{\lambda_{rb}}{(2\pi)^2\varepsilon_{rb}''}\frac{e^2}{\hbar c}\frac{1}{(1-x_F)^3}, \lb{AS}
\eeq
where $\lambda_{rb} = 2\pi c/\omega_{rb}$ corresponds to the red border of photoemission, $\varepsilon_{rb}'' \equiv \varepsilon''(\omega_{rb})$. Coefficient $\eta_s > 1$ when a size of nanoparticle $a<a_s$. Surface photoemission became important, when $\eta_s>1$ so when $a<a_s$, where $a_s$ depends on the frequency $\omega$ of applied field. Otherwise, $a_{s0}$ does not depend on $\omega$, so $a_{s0}$ is the estimation for the maximum size of a nanoparticle (in given environment), when SPE became comparable with VPE. $a_{s0} = 42$~nm for gold spherical nanoparticle in a semiconductor (like GaAs) with parameters as in \ct{C3NR06679G} $V_0=6.31$~eV, $E_F=5.51$~eV, $\varepsilon_{out}=13$, gold dielectric function 
\beq
        \varepsilon_{Au}(\lambda) = 12-\left(\frac{\lambda}{\lambda_p}\right)^2\frac{1}{1+i\lambda/\lambda_c} \lb{gdf}
\eeq
used in \ct{Protsenko_2012}, with $\lambda_p=0.136$~$\mu$m, $\lambda_c = 55$~$\mu$m.
%
%
\begin{figure}[ht]
\bc \includegraphics[width=7cm]{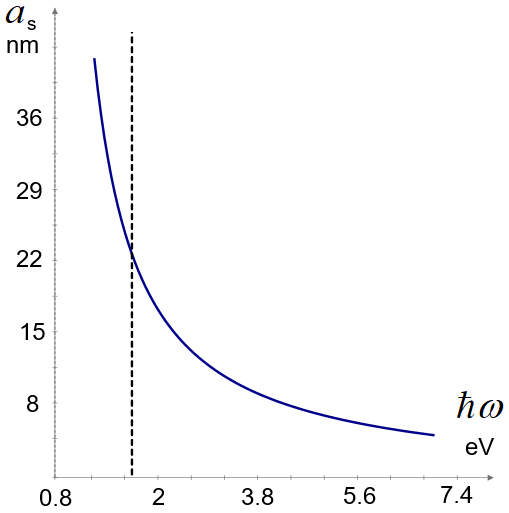} \ec
\vspace{-4mm}
\caption{SPE became compatible with VPE, when the radius of nanoparticle $a<a_s(\omega)$ shown in the figure. Vertical dashed line marks the energy of localized plasmon resonance.}\label{Fig2}
\end{figure}
%
%
Fig.~\ref{Fig2} shows $a_s(\hbar\omega)$ for such parameters.
\subsection{SPE and SPA with rectangular-step potential barrier}
Wave functions $\Psi_{0,\pm}$ for rectangular potential barrier $V=0$,  $z<0$, $V=V_0$,  $z>0$
are well-known and given, for example, in \ct{Protsenko_2012}. With such wave functions we find $(\Psi_0\Psi_{1\pm})_{z=0}$, $[\Psi_0'\Psi_{1\pm}'/m^2]_{z=0}$, see Appendix, and obtain 
\beq
C_{\pm}^{(0)}=\frac{|e|}{im_{in}\omega^2}u(k_z)K^{dis}_{\pm}(k_z,k_{\parallel}),\lb{rect_C}
\eeq
with
\beq
u=\frac{k_z}{\left\{k_z+i\sqrt{r_m[k_V^2+k_{\parallel}^2(r_m-1)-k_z^2]}\right\}\left\{\sqrt{k_z^2+k_{\omega}^2}+\sqrt{r_m\left[k_z^2+k_{\omega}^2-k_V^2-k_{\parallel}^2(r_m-1)\right]}\right\}}\lb{s1_2}
\eeq
and
\[
K^{dis}_{\pm} = \left(r_{\varepsilon}r_m- 1\right)\left(k_z^2+\frac{k_{\omega}^2}{2}\right) - r_{\varepsilon}r_m\left[k_V^2+k_{\parallel}^2\left(r_m-1\right)\right] \pm i(r_{\varepsilon}-1) K_{\pm}\sqrt{r_m[k_V^2+k_{\parallel}^2(r_m-1)-k_z^2]}, 
\]\beq
K_+ = \sqrt{k_z^2+k_{\omega}^2}, \hspace{0.5cm} K_- = r_m\sqrt{k_z^2+k_{\omega}^2-k_V^2-k_{\parallel}^2(r_m-1)}. \lb{s1_1}
\eeq
We take dimensionless   variables
\beq
x_z=(k_z/k_V)^2, \hspace{0.5cm} x_{\parallel}=(k_{\parallel}/k_V)^2, \hspace{0.5cm} x_{\omega} = (k_{\omega}/k_V)^2 \lb{dimvar1}
\eeq
and  express
\beqr
    \eta_{SPE} &=& \frac{a_s\sqrt{r_m}}{a}\int_0^{\infty}dx_{\parallel}\int_0^{\infty}\frac{dx_z}{\sqrt{x_z}}\Re{\left[\sqrt{x_z+x_{\omega}-1-x_{\parallel}(r_m-1)}\right]}|\tilde{u}\tilde{K}^{dis}_+|^2\tilde{f}_F(x)[1-\tilde{f}_F(x+x_{\omega})],\nonumber\\
    \eta_{SPA} &=& \frac{a_s}{a}\int_0^{\infty}dx_{\parallel}\int_0^{\infty}\frac{dx_z}{\sqrt{x_z}}\sqrt{x_z+x_{\omega}}|\tilde{u}\tilde{K}^{dis}_-|^2\tilde{f}_F(x)[1-\tilde{f}_F(x+x_{\omega})],\nonumber
\eeqr
where   $x=x_z+x_{\parallel}$ is normalizes energy of an electron, functions with tildes depend on normalized variables, as \rf{dimvar1}, so that
\beq
\tilde{u}=\frac{\sqrt{x_z}}{\left\{\sqrt{x_z}+i\sqrt{r_m[1+x_{\parallel}(r_m-1)-x_z]}\right\}\left\{\sqrt{x_z+x_{\omega}}+\sqrt{r_m\left[x_z+x_{\omega}-1-x_{\parallel}(r_m-1)\right]}\right\}}\lb{s1_2_1}.
\eeq
\[
\tilde{K}^{dis}_{\pm} = \left(r_{\varepsilon}r_m- 1\right)\left(x_z+\frac{x_{\omega}}{2}\right) - r_{\varepsilon}r_m\left[1+x_{\parallel}\left(r_m-1\right)\right] \pm i(r_{\varepsilon}-1) \tilde{K}_{\pm}\sqrt{r_m[1+x_{\parallel}(r_m-1)-x_z]}, 
\]\beq
\tilde{K}_+ = \sqrt{x_z+x_{\omega}}, \hspace{0.5cm} \tilde{K}_- = r_m\sqrt{x_z+x_{\omega}-1-x_{\parallel}(r_m-1)}. \lb{s1_1_2}
\eeq
By setting $r_m = 1$ we obtain  $\tilde{K}^{dis}_+ = -K_{\Delta\varepsilon}$, where $K_{\Delta\varepsilon}$ given by Eq.~(17) of \ct{C3NR06679G} and $|\tilde{u}(x_z)|^2$ is the same as $G(x_z)$ given by Eq.~(16) of \ct{C3NR06679G}.

\begin{figure}[ht]
\bc \includegraphics[width=9cm]{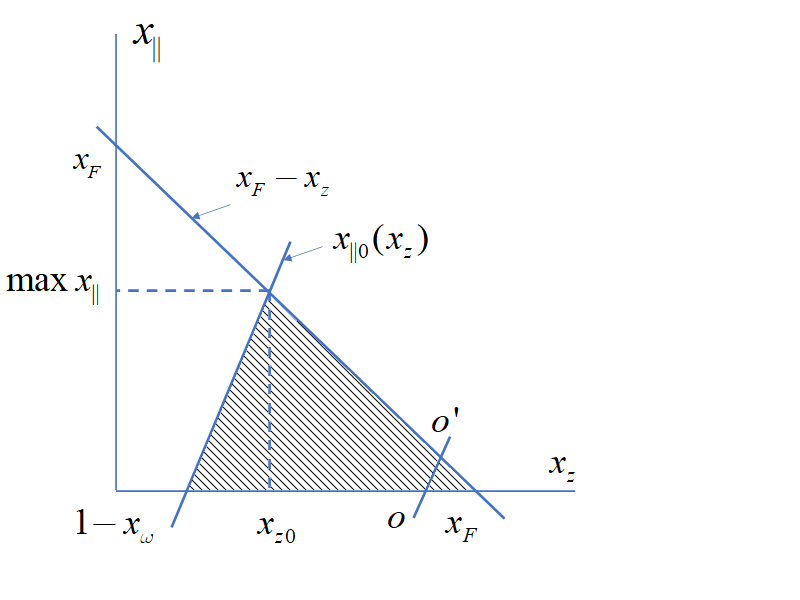} \ec
\vspace{-4mm}
\caption{Area of integration in Eq.~\rf{zero_T_SPE} is shadowed, $\max{x_{\parallel}} = (x_F+x_{\omega}-1)/r_m$. The line $oo'$ restrict integration area at small excess of photon energy $x_{\omega}+x_F-1 \ll x_F$ above the red border, then  $x_z\approx x_F$ and $x_{\parallel} \approx 0$. }\label{Fig3}
\end{figure}

Below we will neglect by the temperature dependence of Ferm-distribution \rf{Fermi_df_1}, set there $T=0$, then  
\beq
    \hspace{-1cm}\eta_{SPE} = \frac{a_s\sqrt{r_m}}{a}\left(\int_{1-x_{\omega}}^{x_{z0}}dx_z\int_0^{x_{\parallel 0}(x_z)}dx_{\parallel}+\int_{x_{z0}}^{x_F}dx_z\int_0^{x_F-x_z}dx_{\parallel}\right)\frac{|\tilde{u}\tilde{K}^{dis}_+|^2}{\sqrt{x_z}}\Re{\sqrt{x_z+x_{\omega}-1-x_{\parallel}(r_m-1)}}, \lb{zero_T_SPE}
\eeq
where integration limits
\[
    x_{z0} = x_F+(1-x_{\omega}-x_F)/r_m, \hspace{0.5cm} x_{\parallel 0}(x_z) = (x_z+x_{\omega} - 1)/(r_m-1), 
\]
the area of integration  on $(x_z,x_{\parallel})$ plane in Eq.~\rf{zero_T_SPE} is shown in Fig.~\ref{Fig3}.  
\subsection{Surface  photoemission near the red border}
\begin{figure}[ht]
\bc \includegraphics[width=6cm]{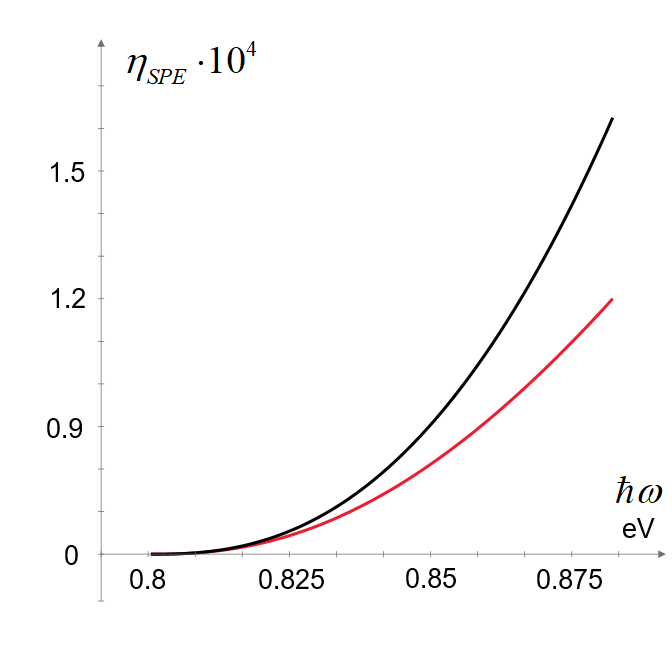} \ec
\vspace{-4mm}
\caption{$\eta_{SPE}$ for rectangular potential barrier. Exact curve Eq.~\rf{zero_T_SPE} (red) and its approximation \rf{ETA_SPE_app_1} near the red border (black).}\label{Fig4}
\end{figure}

Near the red border of photoemission $x_z\approx x_F$, $x_{\parallel}\approx 0$, and $x_z+x_{\omega} \approx 1$, see Fig.~\ref{Fig4}, and we approximate
\beq
    \tilde{u} \approx \frac{1}{1+i\sqrt{r_m (1-x_F)/x_F}}, \hspace{0.5cm} \tilde{K}^{dis}_+ \approx (r_{\varepsilon}r_m - 1)(x_F+1)/2 - r_{\varepsilon}r_m + i(r_{\varepsilon}-1)\sqrt{r_m(1-x_F)}, \lb{app_SPE}
\eeq
and $\sqrt{x_z} \approx \sqrt{x_F}$. We  use approximations \rf{app_SPE}, take ${|\tilde{u}\tilde{K}^{dis}_+|^2}/{\sqrt{x_z}}$ out of the integral in Eq.~\rf{zero_T_SPE}, 
%
%
%
calculate
\beq
    \int_0^ydx_{\parallel}\sqrt{x_z+x_{\omega} - 1 - x_{\parallel}(r_m-1)} = \frac{2}{3(r_m-1)}\left\{(x_z+x_{\omega}-1)^{3/2} - [x_z+x_{\omega}-1-(r_m-1)y]^{3/2}\right\}, \lb{int_aux}
\eeq
calculate the rest of integrals in Eq.~\rf{zero_T_SPE} and obtain
\beq
         \eta_{SPE} = \frac{a_{s0}}{a}\frac{4\sqrt{x_F}}{15}F_{\varepsilon m}\frac{\delta x_{\omega}^{5/2}}{\sqrt{r_m}}, \hspace{0.5cm}F_{\varepsilon m}=\frac{[1+x_F+r_{\varepsilon}r_m(1-x_F)]^2/4+(r_{\varepsilon}-1)^2r_m(1-x_F)}{x_F+r_m(1-x_F)}.  \lb{ETA_SPE_app_1}
\eeq
Here $\delta x_{\omega} = x_{\omega} + x_F-1 \ll 1$ is a part of photon energy above the barrier, $F_{\varepsilon m}=1$ when $r_m=r_{\varepsilon}=1$ and near the red border we take $a_s=a_{s0}$. 
%
%
%
Fig.~\ref{Fig4} shows $\eta_{SPE}(\omega)$ given by Eq.~\rf{zero_T_SPE} and its approximation near the red border \rf{ETA_SPE_app_1} for gold nanoparticle of radius $a= 8.4$~nm in semiconductor like GaAs with $\varepsilon_{out} = 13$ and $r_m=2$. One can see that SPE spectrum $\eta_{SPE}\sim \delta x_{\omega}^{5/2}$, which is good approximation in small region of $\hbar\omega$ near the red border.
\section{Volume photo-emission}\label{Sec3}
We consider the volume photo-emission (VPE) for small spherical nanoparticles  following \ct{C3NR06679G}. We suppose, that only  electrons of metal absorb photons. Then the rate of absorption  $r_{abs}$ given by Eq.~\rf{rabs_v} is also the rate of generation of hot electrons in the unit of volume. 
The rate of  photoemission of hot electrons from the unit of volume  is
\beq
r_{VPE} = r_{abs}W_tW_p, \lb{W_VPE}
\eeq
where $W_t$ is the probability that hot electron avoids inelastic collisions and reaches the nanoparticle surface. $W_p$ is the probability that hot electrons passes through the barrier on the surface and leaves the nanoparticle.  The rate of photoemission from the nanoparticle is
\beq
        R_{VPE} = \int_{V_p}r_{abs}\left<W_tW_p\right>dV, \lb{RVPE_0}
\eeq
where
\[
\left<W_tW_p\right> = \frac{1}{n_h}\int W_tW_pf_hgd^3k 
\]
is averaged over parameters of hot electrons with distribution function $f_h$,  density of states  $g=2/(2\pi)^3$ in the unit of volume, $d^3k = k^2dk\sin{\theta}d\theta d\varphi$ and where $n_h = \int f_h gd^3k$ is the number of hot electron states in the unit of volume. Internal quantum efficiency of VPE is
\beq
    \eta_{VPE} = R_{VPE}/R_{abs} \lb{eta_VPE_0}
\eeq
where $R_{abs}$ is the rate \rf{RABS} of absorption of photons in the nanoparticle. We consider small spherical nanoparticles, where the energy density of electric field $|E_{in}|^2$ is constant. Combyning Eqs~\rf{RVPE_0}, \rf{eta_VPE_0} and \rf{RABS} we express
\beq
\eta_{VPE} = \frac{1}{n_hV_p}\int W_tW_pf_hgd^3kdV. \lb{eta_VPE}
\eeq
Parameters of hot electrons in any point of the volume of small nanoparticle are same, therefore $W_pf_hg$ in Eq.~\rf{eta_VPE} does not depend on spatial coordinates, so we simplify Eq.~\rf{eta_VPE} by separating  there integrations 
\beq
\eta_{VPE} = \frac{1}{n_h}\int \overline{W}_tW_pf_hgd^3k, \hspace{0.5cm} \overline{W}_t = \frac{1}{V_p}\int W_tdV. \lb{eta_VPE_1a}
\eeq
Following \ct{PhysRev.38.45,doi:10.1002/pip.2278} we approximate the distribution of hot electrons by homogeneous distribution over the energies from $E_F$ to $E_F+\hbar\omega$ so 
  $f_h = 1$ for $k_F^2<k^2<k_F^2+k_{\omega}^2$ and $f_h=0$ otherwise. 
%
%
\begin{figure}[ht]
\bc \includegraphics[width=5cm]{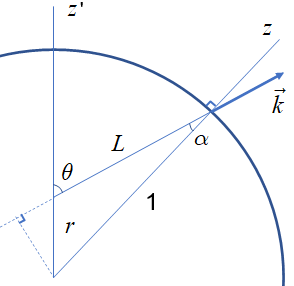} \ec
\vspace{-4mm}
\caption{Spherical coordinate system in $k$-space for volume photoemission with z-axes $z'$ and polar angle $\theta$. Axes $z$ is normal to the nanoparticle surface in the point, where  electron leaves  nanoparticle. The distance $r$ is from the center of nanoparticle to the point of generation of hot electron. All distances are normalized to nanoparticle radius  taken 1 in the Figure.}\label{Fig5}
\end{figure}
%
%
Nothing depends on angular coordinates in space and on ashimulal angle in $k$-space in spherical nanoparticle, so the integration over these variables in Eq.~\rf{eta_VPE} gives $4\pi$ and $2\pi$, respectively. We calculate 
\[
    n_h = 4\pi\int_{k_F}^{k_h}\frac{2}{(2\pi)^3}k^2dk = \frac{k_h^3-k_F^3}{3\pi^2}, \hspace{0.5cm} V_p=4\pi/3,\hspace{0.5cm} k_h^2=k_F^2+k_{\omega}^2 
\]
and write integratls in Eq.~\rf{eta_VPE}  in coordinates shown in Fig.~\ref{Fig5}
\beq
\eta_{VPE} = \frac{9}{2(k_h^3-k_F^3)}\int_0^1r^2dr\int_{k_V}^{k_h}k^2dk\int_0^{\pi}\sin{\theta}d\theta W_t(r,\theta)\Re{[W_p(k,\theta)]}. \lb{eta_VPE_1}
\eeq
Here and below we use dimensionless spatial coordinate $r$ normalised to nanoparticle radius. Taking $\Re{[W_p(k,\theta)]}$ in Eq.~\rf{eta_VPE_1} we consider there only hot electrons passing the barrier, such electrons fly in the cone with $0<\alpha < \alpha_m$ (see $\alpha$ in Fig.\ref{Fig5})   \ct{Khurgin}. For such electrons $W_p(k,\theta)$ is real, otherwise $W_p(k,\theta)$ is purely imaginary. We take dimensionless parameters 
\beq
x=E/V_0, \hspace{0.5cm} x_F=E_F/V_0, \hspace{0.5cm} x_h=(E_F+\hbar\omega)/V_0  \lb{dimvar_VPE}
\eeq
and re-write Eq.~\rf{eta_VPE_1} in terms of them. In general $W_p(k,\theta) \neq W_p(x,\theta)$ but, for simplicity, we keep the same notations for $W_p(k,\theta)$ and $W_p(x,\theta)$ as well as for another similar notations,
\beq
\eta_{VPE} = \frac{9}{4(x_h^{3/2}-x_F^{3/2})}\int_0^1r^2dr\int_{1}^{x_h}\sqrt{x}dx\int_0^{\pi}\sin{\theta}d\theta {W}_t(r,\theta)\Re{[{W}_p(x,\theta)]}. \lb{eta_VPE_2}
\eeq
Following \ct{C3NR06679G} we suppose that hot electron moves ballistically without elastic collisions, and it can not  be emitted if it looses the energy in, at least, single inelastic collision. Then the probability that hot electron reaches the nanoparticle boundary is
\beq
        W_t(r,\theta) = \exp{[-L(r,\theta)/l_e]}, \lb{WTR}
\eeq
where $L(r,\theta) = \sqrt{1-r^2\sin^2{\theta}} - r\cos{\theta}$ is the length of the way of hot electron to the interface shown in Fig.~\ref{Fig5}, $r$ is the distance from the nanoparticle center to the point of generation of hot electron,  and $l_e$ is the mean free path of an electron in metal, $r$ and $l_e$ are normalized to the radius of spherical nanoparticle.

We simplify Eq.~\rf{eta_VPE_2} and represent it according with Eq.~\rf{eta_VPE_1a}. First, we separate the integration  over $d\theta$ in Eq.\rf{eta_VPE_2} in two parts: from $0$ to $\pi/2$ and from $\pi/2$ to $\pi$. Probability ${W}_p$ that hot electron passes the barrier depends on normalised kinetic energy $x$ of the electron and on the part $x\sin^2{\alpha}$ of this energy, correspondent to the motion parallel to the interface, therefore ${W}_p = W_p(x,\sin^2{\alpha})$. We note that $\sin^2{\alpha} = r^2\sin^2{\theta}$ (see Fig.~\ref{Fig5}) and introduce in Eq.~\rf{eta_VPE_2} new variable $y = r^2\sin^2{\theta}$ instead of $\theta$, so  ${W}_p={W}_p(x,y)$.  We come from $\theta$ to $y$ in Eqs.~\rf{eta_VPE_2}, \rf{WTR}, taking into account that $\sin{\theta}d\theta = dy/(2r\sqrt{r^2-y})$, introduce $z=r^2$ and separate the integration in $\eta_{VPE}$ as in Eq.~\rf{eta_VPE_1a}
\beq
\eta_{VPE} = \frac{9}{4(x_h^{3/2}-x_F^{3/2})}\int_{1}^{x_h}\sqrt{x}dx\int_0^{y_m(x)}\overline{W}_t(y){W}_p(x,y)dy, \lb{eta_VPE_49}
\eeq
where
\beq
\overline{W}_t(y) = \frac{1}{4}\int_y^1\frac{dz}{\sqrt{z-y}}W_t(z,y)\lb{Wbar}
\eeq
with 
\beq
W_t=\exp{\left(-\sqrt{1-y}/l_e\right)}\left[\exp{\left(-\sqrt{z-y}/l_e\right)}+\exp{\left(\sqrt{z-y}/l_e\right)}\right], \lb{int_0}
\eeq
where $W_t$ is given initially  by Eq.~\rf{WTR}. The integral \rf{Wbar} can be taken and we obtain
 \beq
 \overline{W}_t(y) = \frac{l_e}{2}\left[1-\exp\left(-\frac{2\sqrt{1-y}}{l_e}\right)\right].\lb{Wint}
 \eeq
Now we determine $y_m(x)$ in Eq.~\rf{eta_VPE_49}. The energy of hot electron in metal is  $(\hbar k)^2/2m_{in}$. When the electron passes through the barrier, the component $k\sin{\alpha}$ of its wave vector is preserved, while the component ${k}_z$ normal to the interface is not preserved. Z-component $\tilde{k}_z$ of the wave vector of emitted electron outside the metal, far from the barrier,  is found from the energy conservation law
\beq
    {\hbar^2(\tilde{k}_z^2+k^2\sin^2{\alpha}) }/{2m_{out}}  + V_0 =  (\hbar k)^2/{2m_{in}}. \lb{disp_VPE}
\eeq
Taking $\sin{\alpha}=r\sin{\theta}$, see  Fig.~\ref{Fig5}, we obtain from Eq.~\rf{disp_VPE}
\beq
\tilde{k}_z = \sqrt{[k^2(1-r_mr^2\sin^2{\theta}) - k_V^2]/r_m}. \lb{k_zv}
\eeq
$\tilde{k}_z$ must be real, so only electrons with $k>k_V$ and inside a cone with $\sin{\theta} <\sin{\theta_{m}}$ pass through the barrier. From Eq.~\rf{k_zv} we obtain
\beq
r^2\sin^2{\theta}<r^2\sin^2{\theta_{m}}\equiv y_m(x)=(1-1/x)/r_m, \hspace{0.5cm} x = k^2/k_V^2 \lb{cone_theta}
\eeq
 We suppose, for simplicity, $r_m>1$ which is typical case for semiconductir-metal interface, then $0<y_m(x)<1$. With the result \rf{Wint} we obtain
\beq
\eta_{VPE} = \frac{9l_e}{8(x_h^{3/2}-x_F^{3/2})}\int_{1}^{x_h}\sqrt{x}dx\int_0^{(1-1/x)/r_m}\left[1-\exp\left(-\frac{2\sqrt{1-y}}{l_e}\right)\right]{W}_p(x,y)dy. \lb{eta_VPE_50}
\eeq
Since we do not know $l_e$ precisely, we can ignore y-dependence in $\exp\left(-{\sqrt{1-y}}/{l_e}\right)$ in Eq.~\rf{eta_VPE_50} and write in good approximation
\beq
\eta_{VPE} = \frac{9l_e[1-\exp{(-2/l_e)}]}{8(x_h^{3/2}-x_F^{3/2})}\int_{1}^{x_h}\sqrt{x}dx\int_0^{(1-1/x)/r_m}{W}_p(x,y)dy. \lb{eta_VPE_51}
\eeq
This result can be used with arbitrary ${W}_p(x,y)$ with $W_t$ given by Eq.~\rf{WTR}.
\subsection{VPE with rectangular-step potential barrier}
We calculate the wave function of hot electron in the  coordinate system with axes $z$, normal to the nanoparticle surface, shown in  Fig.~\ref{Fig5}, as we did for SPE and SPA. Similar with the wave-functions \rf{Fac_wf} for SPE, the wave function of hot electron in VPE is factorized as $\Psi_z(z)e^{i\vec{k}_{\parallel}\vec{\rho}}$. In Fig.\ref{Fig5} we see that 
\beq
{k}_{\parallel} = k\sin{\alpha}, \hspace{0.5cm} {k}_z = k\cos{\alpha} \lb{sin_cos}
\eeq
are components of the wave-vector of hot electron, parallel and perpendicular to the surface, the wave number of hot electron is $k$. Z-dependent part  of the wave function is 
\beq
\Psi_z(z) = \left(e^{ik_zz}+Ae^{-ik_zz}\right)_{z<0} + (Be^{i\tilde{k}_zz})_{z>0}, \lb{wf_step_b}
\eeq
where $A$ and $B$ are c-number constants. $z$-component $\tilde{k}_z$ of the wave-vector for $z>0$ is given by Eq.~\rf{k_zv}. Inserting the wave function \rf{wf_step_b} into the boundary conditions at $z=0$ $\Psi_h(-0) = \Psi_h(+0)$ and $m_{in}^{-1}(d\Psi_h/dz)_{z=-0} =m_{out}^{-1}(d\Psi_h/dz)_{z=+0}$ we obtain $B=2/(1 + r_m\tilde{k}_z/k_z)$. With the wave-function \rf{wf_step_b} the fluxes of electrons in z-direction toward (away from) the barrier are $j^{(in)}_z = \hbar k_z/m_{in}$ ($j^{(out)}_z = (\hbar\tilde{k}_z/m_{out})|B|^2$), so  the probability that hot electron  passes through the barrier is
\beq
    W_p(k,k_z) \equiv {j^{(out)}_z}/{j^{(in)}_z} = \frac{ 4r_mk_z\Re{\tilde{k}_z}}{(k_z+r_m\tilde{k}_z)^2}, \lb{Wpass}
\eeq
if conditions \rf{cone_theta} are true and $W_p =0$ otherwise, $k_F=\sqrt{2m_{in}E_F}/\hbar$, $k_{\omega}=\sqrt{2m_{in}\omega/\hbar}$.  Using dimensionless variables \rf{dimvar_VPE} and $y=r^2\sin^2{\theta}$ we write
\beq
    {W}_p(x,y)=\frac{4\sqrt{r_mx(1-y)[x(1-r_my)-1]}}{\left\{\sqrt{x(1-y)}+\sqrt{r_m[x(1-r_my)-1]}\right\}^2}. \lb{Wpass_dim}
\eeq
Volume photoemission efficiency $\eta_{VPE}$ is given by \rf{eta_VPE_51}.
\subsection{VPE near the red border}
Suppose, the absorption of photons leads to a small excess of the hot electron energy above the barrier $\delta x_{\omega} =(\hbar\omega + E_F)/V_0-1 \ll 1$. Then $x=1+\delta{x}$, $\delta{x} \ll 1$, $y\ll 1$ and we approximate $W_p$ in Eq.~\rf{Wpass_dim} as
\beq
    {W}_p \approx 
    4\sqrt{r_m}\sqrt{ \delta{x}_{\omega}-r_my} \lb{border}
\eeq
Inserting Eq.~\rf{border} into Eq.~\rf{eta_VPE_51} we take  integrals we obtain
\beq
  \eta_{VPE} \approx \frac{6}{5}\frac{l_e[1-\exp{(-2/l_e)}]}{1-x_F^{3/2}}\cdot\frac{\delta x_{\omega}^{5/2}}{\sqrt{r_m}}. \lb{eta_VPE_4}  
\eeq
We see that near the red border of photoemission $\eta_{VPE}\sim 1/\sqrt{r_m}$. Such dependence of $\eta_{VPE}$ on the electron effective mass is weaker than the result (26) of \ct{Khurgin} where $\eta_{VPE} \sim 1/r_m$ near the red border. This difference is because of in Eq.~(26) of \ct{Khurgin}  $R_{eff}(0)$ (the analog of our $W_p \sim \sqrt{r_m}$) does not depend on $r_m$. The change in the effective mass (for $r_m>1$) effectively increases the potential barrier on the interface, because the momentum conservation law. However the same acceleration increases the flux of electrons in the normal direction out of the surface and therefore increases the photoemission current, this is why our $W_p\sim \sqrt{r_m}$.  Overall influence of the decreace of effective mass on the photoemission is still negative, because the pass through the barrier requires the increase of the electron kinetic energy proportional to the square of the electron velocity. 

Near the red border, the spectra of SPE and VPE are both $\sim \delta x_{\omega}^{5/2}$. However approximations \rf{ETA_SPE_app_1} and \rf{border} are valid only in a small interval of frequencies near the red border. Analytical approximation for $\eta_{VPE}$ valid in lager region of frequencies of applied field is obtained by taking 
\beq
        W_p \approx \frac{4\sqrt{r_m}\sqrt{ \delta{x}_{\omega}-r_my}}{(1+\sqrt{r_m}\sqrt{ \delta{x}_{\omega}-r_my})^2}. \lb{WP_app2}
\eeq  
and $\sqrt{x} \approx 1$ in Eq.~\rf{eta_VPE_51}. Then integrals in Eq.~\rf{eta_VPE_51} can be calculated and we find
\beq
        \eta_{VPE} = \frac{18l_e(1-e^{-2/l_e})}{r_m^2(1-x_F^{3/2})}\left[\delta x_{\omega}(1+\sqrt{r_m\delta x_{\omega}}/3)-2\sqrt{\delta x_{\omega}/r_m}+\ln{(\sqrt{r_m\delta x_{\omega}}+1)}(2/r_m-\delta x_{\omega})\right]. \lb{app_bet}
\eeq
Comparing approximations \rf{border} and \rf{app_bet} we see that dependencies $\eta_{VPE}(\delta x_{\omega}, r_m)$ not very close to the red border strongly depend on the expression for the  probability $W_p$ of the event, that hot electron passes the barrier. The same is true for SPE, where we do not yet find   approximation similar to \rf{app_bet}. 

\begin{figure}[ht]
\bc \includegraphics[width=7cm]{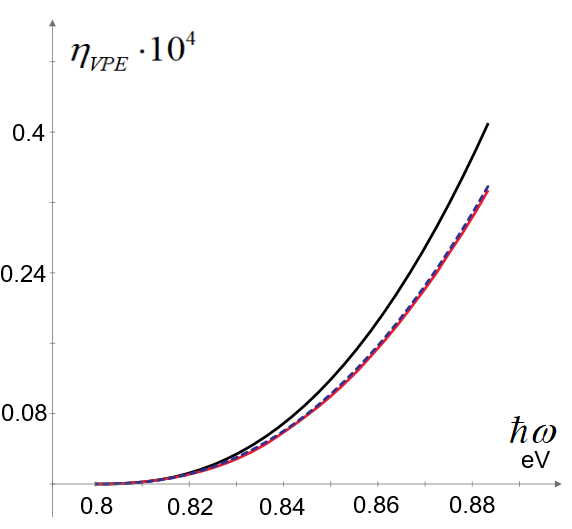} \ec
\vspace{-4mm}
\caption{Internal quantum efficiency of volume photoemission \rf{eta_VPE_51} with exact rectangular step barrier \rf{Wpass_dim} (red curve), its approximation near the red border \rf{eta_VPE_4} (black curve) and with more precise approximation \rf{app_bet} (blue dashed curve).}\label{Fig6}
\end{figure}

Fig.~\ref{Fig6} shows $\eta_{VPE}$  given by Eq.~\rf{eta_VPE_50}, its approximation \rf{border} near the red border and more precise approximation \rf{app_bet} for $l_e = 0.5$ (in units of nanoparticle radius), $r_m=2$ for gold spherical nanoparticle.
%
%
%
%
\section{Comparison of efficiencies of SPE and VPE}\label{Sec4}
Comparing  Fig.~\ref{Fig4} with Fig.~\ref{Fig6} we  see that near the red border $\eta_{SPE} > \eta_{VPE}$, so SPE is more efficient that VPE  for gold nanoparticle of diameter less than $2a=15$~nm. Equating Eqs.~\rf{ETA_SPE_app_1} and \rf{eta_VPE_4} we find the maximum nanoparticle radius $a_{rb}(r_m,r_{\varepsilon})$, when efficiencies $\eta_{VPE} = \eta_{SPE}$ on the red border
\beq
        a_{rb}(r_m,r_{\varepsilon}) = a_{s0}\frac{2\sqrt{x_F}}{9}F(r_m,r_{\varepsilon})\frac{1-x_F^{3/2}}{l_e[1-\exp{(-2/l_e)}]}, \lb{a_rb}
\eeq
where $F(r_m,r_{\varepsilon})$ is given by Eq.~\rf{ETA_SPE_app_1}. Therefore SPE will be more efficient that VPE on the red border for spherical gold nanoparticles of radius $a<a_{rb}$ calculated for given semiconductor environment.
%
%
\begin{figure}[ht]
\bc \includegraphics[width=7cm]{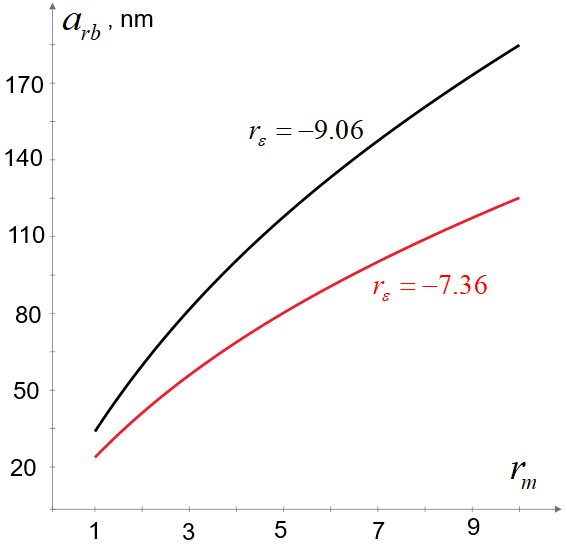} \ec
\vspace{-4mm}
\caption{Nanoparticle radius $a_{rb}$, when $\eta_{SPE} = \eta_{VPE}$ so SPE and VPE have the same efficiencies on the red border of photoemission. SPE is more efficient that VPE on the red border for nanoparticles of radius $a<a_{rb}$. Black curve is for gold nanoparticle in the semiconductor with $\varepsilon_{out} = 13$, red curve is for $\varepsilon_{out} = 16$. $a_{rb}$ increases with $|r_{\varepsilon}|$.   }\label{Fig7}
\end{figure}
%
%

Fig~\ref{Fig7} shows $a_{rb}(r_m)$ for various $r_{\varepsilon}$. Note in Fig~\ref{Fig7} that larder jump in the  dielectric function $|r_{\varepsilon}| = |\varepsilon_{in}/\varepsilon_{out}|$ corresponds larder nanoparticle radius $a_{rb}$ when $\eta_{SPE} = \eta_{VPE}$. 

\begin{figure}[ht]
\bc \includegraphics[width=7cm]{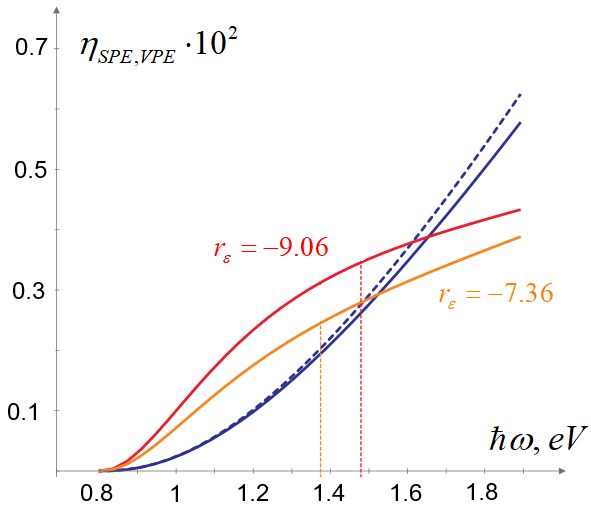} \ec
\vspace{-4mm}
\caption{Spectra of internal quantum efficiencies of VPE (blue curves) and SPE (red and orange curves) given by Eqs.~\rf{eta_VPE_51} and \rf{zero_T_SPE}, respectively. Blue dashed curve is approximation \rf{app_bet} for $\eta_{VPE}$. Vertical dotted lines mark plasmon resonance frequencies. The red curve is for gold nanoparticle in $GaAs$-like semiconductor, the red curve is for silica environment. The change in the electron effective mass on the interface is $r_m=2$, diamerer of spherical nanoparticle $2a=15$~nm, the electron mean free pass $l_e=0.5a$. $\eta_{SPE}>\eta_{VPE}$ if photon frequency is not too high, including plasmon resonance frequencies. }\label{Fig8}
\end{figure}

Fig~\ref{Fig8} shows spectra of $\eta_{SPE}(\omega)$~\rf{zero_T_SPE} and 
$\eta_{VPE}(\omega)$~\rf{eta_VPE_51} for $\varepsilon_{out} = 13$ and $\varepsilon_{out} = 16$
for $r_m=2$ and nanoparticle of diameter $2a=15$~nm. We see a region, with not too high  photon energy, where   $\eta_{SPE}>\eta_{VPE}$, so SPE is more efficient than VPE. The region with $\eta_{SPE}>\eta_{VPE}$   
includes  plasmon resonance frequency. 
%
%
%
%

\begin{figure}[ht]
\bc \includegraphics[width=7cm]{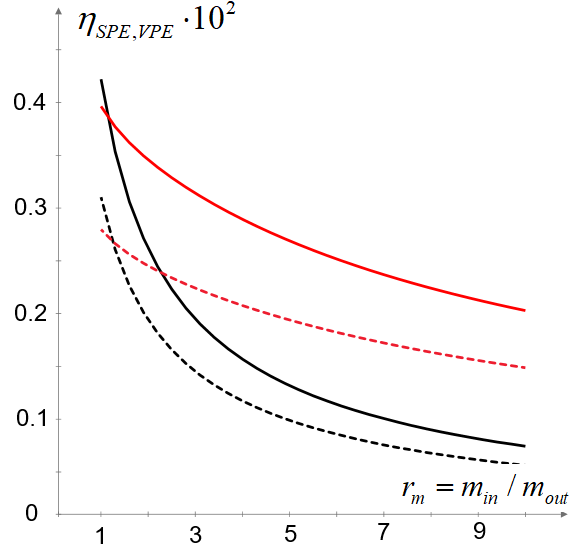} \ec
\vspace{-4mm}
\caption{Quantum efficiencies of SPE (red curves) and VPE (black curves) at plasmon resonance frequency as functions of the electron effective mass change $r_m$ on the metal-semiconductor interface. Solid curves are for gold nanoparticle in $GaAs$-like semiconductor, dashed curves correspond to silica semiconductor environment. The difference in $\eta_{VPE}$ (black curves) in different environment is because of different plasmon resonance frequencies. $\eta_{SPE} > \eta_{VPE}$ everywhere apart of small region near $r_m=1$. }\label{Fig9}
\end{figure}

Fig.~\ref{Fig9} shows $\eta_{SPE}$ and $\eta_{VPE}$ as functions of  the jump in the electron effective mass $r_m$ for gold spherical nanoparticle in $GaAs$-like semiconductor with $\varepsilon_{out} = 13$ and in silica environment with $\varepsilon_{out} = 16$ at plasmon resonance frequency (different in different  environments).  Both SPE and VPE efficiencies $\eta_{SPE}$ and  $\eta_{VPE}$ decrease with  $r_m$, but $\eta_{SPE}$ decreases slowly than $\eta_{VPE}$. Because of that SPE is more efficient: $\eta_{SPE}>\eta_{VPE}$ everywhere, apart of small region of $r_m$ near $r_m=1$. 
%
%
%
%

In Figs.~\ref{Fig8}, \ref{Fig9} and \ref{Fig7} we see that SPE efficiency is larger for larger break $|r_{\varepsilon}|$ in the  electromagnetic field (EMF) on the interface. This is because of the absorption of photons by electrons, interacting with EMF discontinuity, is added to the absorption at collisions of electrons with potential barrier on the interface \ct{C3NR06679G} at SPE. Absorption at EMF discontinuity is the process  reverse  to well-known transition radiation, when a charged particle passes through inhomogeneous media, for example, through a boundary between two different media \ct{Tr_rad}.

Slow decrease of $\eta_{SPE}$ with $r_m$, respectively to  fast decrease of $\eta_{VPE}$,  is because of the electron wave vector component $k_{\parallel}$, preserved on the interface, remains small, the same as for cold electron, in case of SPE. In other words, hot electrons must be accelerated, due to the momentum conservarion law, passing the interface at VPE, while in case of SPE such acceleration is required for "cold" electrons. So less energy is needed for the acceleration in the case of SPE that in VPE. The electron at SPE interacts only with  normal component of EMF, and the energy of absorbed photon comes only to the motion of electron normal to the interface. Such motion is affected by the potential barrier, so it does not obey by the momentum conservation law and therefore it is not accelerated due to $m_{in} > m_{out}$. By contrast, in the case of VPE the energy of absorbed photon  is, in average,  equally distributed among all directions of motion: 1/3 of that energy comes to the direction normal to the interface and 2/3 goes to directions parallel to the interface. In case of VPE the motion parallel to the interface requires acceleration (if $r_m>1$). So in order to preserve the momentum parallel to the interface at mass discontinuity, cold electron must have more energy in case of VPE that in case of SPE. 

We express $x_z = x_0\cos^2{\alpha}$, $x_{\parallel} = x_0\sin^2{\alpha}$ and $x=x_0+x_{\omega}$, where $x_0 = E_0/V_0$ is normalized energy of cold electron before absorption of photon (in the volume or on the surface), $x_{\omega} = \hbar\omega/V_0$ and $\alpha$ is the angle of incidence of an electron on the nanopartical surface as shown in Fig.~\ref{Fig5}.  Then we rewrite conditions \rf{SPE_cond} for SPE and \rf{cone_theta} (with $r^2\sin^2{\theta} = \sin^2{\alpha}$) for VPE as, respectively 
\beq
        \sin^2{\alpha} <\sin^2{\alpha_{SPE}}= \delta x_{\omega}/(r_mx_0), \hspace{0.5cm}\sin^2{\alpha}<\sin^2{\alpha_{VPE}} = \delta x_{\omega}/[r_m(x_0+x_{\omega})] \lb{SPE_cond_1}
\eeq
where $\delta x_{\omega} = x_0 + x_{\omega} -1$ is excess of the hot electron energy above the barrier and $\alpha_{SPE,VPE}$ are maximum angles of cones of flying of electrons of VPE and SPE. We see that $\sin^2{\alpha_{SPE}}/\sin^2{\alpha_{VPE}} = 1+x_{\omega}/x_0>1$, so the cone of electrons available for SPE is larger than for VPE. 
\section{Summary}
Decrease of electron effective mass in the semiconductor environment of  metal nanoparticle reduces the surface and the volume photoemission (SPE and VPE)  as a consequence of the momentum conservation law, which requires additional energy from an electron and effectively increases the potential barrier. We find that such reduction for VPE is larder  than for  SPE.  The reason is that the energy of  a photon absorbed at SPE is contributed to the electron motion normal to the interface, the conservation of momentum is not required in this direction. The motion of electron along the interface, subjected by the momentum conservation law, is not accelerated by absorption of photon.  In contrast, the motion of electrons in all directions is  equally accelerated at photon absorption in the volume and VPE. The conservation of momentum parallel to the interface requires larger electron velocity outside the metal and, correspondingly, larger energy of electrons in metal in VPE than in SPE. So that, less electrons are available for photoemission for VPE that for SPE.  We provide general formulas for internal quantum efficiencies of SPE and VPE taking into account  discontinuities in the electron effective mass, dielectric function and potential barrier on the metal-environment interface of metal nanoparticle. We derive analytical formulas for  quantum efficiencies on the red border of photoemission. Using these formulas we compared SPE and VPE from gold spherical nanoparticle in semiconductors with rectangular potential barrier on the interface and find conditions when SPE is more efficient that VPE.  

Results can be used for modelling, investigation and optimisation of conditions  of generation of "hot" electrons at the absorption of electromagnetic field by metal nanoparticles, for calculations of broadening of localized plasmon resonances by collisions of electrons with nanoparticle surface and for heating of nanoparticles by the radiation. 

\section*{Acknowledgments}
A.V.U. acknowledges the Russian Science Foundation (Grant No. 20-19-00559) and I.E.P. thanks Russian Foundation for Basic Research (Grant No. 21-58-15011) for support.

\section*{Appendix}
\subsection*{Derivation of formulas \rf{New1} and \rf{2}}
Following \ct{Protsenko_2012} we write
\beq
C_{\pm} = \frac{|e|m}{W(\hbar\omega)^2}\int_{-\infty}^{\infty}\frac{dz}{m}(c_V+c_{\mathcal{E}}+c_m) \lb{CPM0}
\eeq
where
\beqr
c_V&=&-\mathcal{E}V'\Psi_0\Psi_{1\mp}\nonumber\\
c_{\mathcal{E}} & = & \mathcal{E}'\left[\frac{\hbar^2}{2m}\Psi_0'\Psi_{1\mp}' +\left(E_0-V+\frac{\hbar\omega}{2}\right)\Psi_0\Psi_{1\mp}\right]\lb{c_VEM}\\
c_m & = & -\frac{\mathcal{E}m'}{m}\left(E_0-V+\frac{\hbar\omega}{2}\right)\Psi_0\Psi_{1\mp}. \nonumber
\eeqr
We insert \rf{c_VEM} into \rf{CPM0} and find
\beq
C_{\pm} = \frac{|e|m}{W(\hbar\omega)^2}\int_{-\infty}^{\infty}dz\left\{\left[ -\left(\frac{\mathcal{E}V}{m}\right)' +\left(E_0+\frac{\hbar\omega}{2}\right)\left(\frac{\mathcal{E}}{m}\right)'\right]\Psi_0\Psi_{1\mp}+\mathcal{E}'\frac{\hbar^2\Psi_0'\Psi_{1\mp}'}{2m^2}\right\}. \lb{C_ap}
\eeq
If $f(z)$ is discontinues function in $z=0$, $f(+0) =f_+$, $f(-0) =f_-$ then $f'(z)_{z=0} = (f_{+0} - f_{-0})\delta(0)$. We consider a product of discontinues functions as single discontinues function.  Therefore $(\mathcal{E}V/m)_{z=0}' = (\mathcal{E}_{out}V_0/m_{out})\delta(0)$, taking into account that $V(z<0) = 0$; $(\mathcal{E}/m)_{z=0}' = (\mathcal{E}_{out}/m_{out}-\mathcal{E}_{in}/m_{in})\delta(0)$. From boundary conditions $\varepsilon_{in}\mathcal{E}_{in} = \varepsilon_{out}\mathcal{E}_{out} = \mathcal{E}_n$, therefore $\mathcal{E}_{in}/\mathcal{E}_{out} = \varepsilon_{out}/\varepsilon_{in}$. After integration in \rf{C_ap} we obtain result \rf{New1} with $C_{\pm}^{(0)}$ given by \rf{2}.
\subsection*{Wave functions for rectangular step potential barrier}
For rectangular-step potential \rf{new_10} multipliers $\Psi_{z0,1\pm}$ in wave functions \rf{Fac_wf} are obtained from wave functions $\Psi_{0,1\pm}$ of 1D problem  \ct{Protsenko_2012}  by replacement $V_0\rightarrow V_z$  
\beqr
\Psi_0(z) &=& \left[\exp{(ik_z z)} + A_0\exp{(-ik_z z)}\right]_{z< 0} + B_0\exp{(-\tilde{k}_z} z)_{z> 0},\nonumber\\
\Psi_{1+}(z) &=& [A_{1+}\exp{(ik_{1z}z)} + B_{1+}\exp{(-ik_{1z}z)}]_{z<0}, + \exp{(i\tilde{k}_{1z}z)}_{z>0}\nonumber\\
\Psi_{1-}(z) &=& [ A_{1-}\exp{(i\tilde{k}_{1z}z)} + B_{1-}\exp{ (-i\tilde{k}_{1z}z) }]_{z>0}, + \exp{ (-i{k}_{1z}z) }_{z<0}\nonumber
\eeqr
$\tilde{k}_z =  \sqrt{[  k_V^2+k_{\parallel}^2(r_m-1)-k_z^2]/r_m}$ is real, $k_{1z} = \sqrt{k_z^2+k_{\omega}^2}$, $\tilde{k}_{1z} = \sqrt{[k_z^2+k_{\omega}^2 - k_V^2-k_{\parallel}^2(r_m-1)]/r_m}$, $r_m=m_{in}/m_{out}$ 
\[
    A_0={(1-i\theta_0)}/{(1+i\theta_0)}, \hspace{1cm} B_0=2/(1+i\theta_0),
\]\[
    A_{1-}=(1+\theta_1)/2, \hspace{1cm} B_{1-}=(1-\theta_1)/2,
\]\[
A_{1+}=(\theta_1-1)/2\theta_1, \hspace{1cm} B_{1+}=(1+\theta_1)/2\theta_1
\]
and
\[
\theta_0=\sqrt{r_m\left( {V_z}/{E_z}-1\right)}, \hspace{0.5cm}\theta_1=\sqrt{ r_m[ 1-{V_z}/{(E_z+\hbar\omega)}]}.\]
so that
\[
    (\Psi_0\Psi_{1\pm})_{z=0} = B_0  = \frac{2\sqrt{E_z}}{\sqrt{E_z}+i\sqrt{r_m(V_z-E_z)}}  = \frac{2k_z}{ k_z+i\sqrt{r_m[k_V^2+k_{\parallel}^2(r_m-1)-k_z^2]} }.
\]\[
        \left.\frac{\Psi_0'\Psi_{1-}'}{m^2}\right|_{z=0}=\frac{i\tilde{k}_zk_{1z}}{m_{out}m_{in}}B_0 = 
        i\frac{\sqrt{[  k_V^2+k_{\parallel}^2(r_m-1)-k_z^2](k_{\omega}^2+k_z^2)}}{m_{in}\sqrt{m_{out}m_{in}}}B_0, 
\]\[
        \left.\frac{\Psi_0'\Psi_{1+}'}{m^2}\right|_{z=0}=-\frac{i\tilde{k}_z\tilde{k}_{1z}}{m_{out}^2}B_0 = 
        -i\frac{\sqrt{[ k_V^2+k_{\parallel}^2(r_m-1)-k_z^2][  k_z^2 + k_{\omega}^2 - k_V^2-k_{\parallel}^2(r_m-1)]}}{m_{out}\sqrt{m_{out}m_{in}}}B_0, 
\]and\[
        \frac{W(z)}{m(z)} = \frac{ik_1}{m_{in}}(1+\theta_1) =  \frac{1}{m_{in}}\left\{\sqrt{k_z^2+k_{\omega}^2} +\sqrt{r_m\left[k_z^2+k_{\omega}^2-k_V^2-k_{\parallel}^2(r_m-1)\right]}\right\}
\]

\bibliographystyle{ieeetr}
\bibliography{myrefs.bib}
\end{document}